\documentclass[fleqn,usenatbib,useAMS]{mnras}

\usepackage{graphicx}	
\usepackage{amsmath}	
\usepackage{amssymb}	
\usepackage{multicol}        
\usepackage{bm}		
\usepackage{pdflscape}	
\usepackage{soul}

\newcommand{\Mpc}{\mathrm{Mpc}}
\newcommand{\Gyr}{\mathrm{Gyr}}
\newcommand{\kpc}{\mathrm{kpc}}
\newcommand{\hMpc}{h^{-1} \mathrm{\Mpc}}
\newcommand{\Msol}{\textup{M}_\mathrm{\sun}}
\newcommand{\xMsol}[2]{\ensuremath{{#1}\times 10^{#2} \,\Msol}}
\newcommand{\Mvir}{\mathrm{M}_{200 \mathrm{c}}}
\newcommand{\rvir}{\mathrm{r}_{200 \mathrm{c}}}
\newcommand{\cnfw}{c_{\mathrm{NFW}}}
\newcommand{\spin}{\lambda_{\mathrm{B01}}}
\newcommand{\env}{1 + \delta_{10}}
\newcommand{\zform}{z_{\mathrm{F}}}
\newcommand{\red}[1]{\color{black}{#1}}

\usepackage[T1]{fontenc}
\usepackage{ae,aecompl}

\usepackage{newtxtext,newtxmath}
\usepackage{soul}

\title[Variance modified haloes]
{Sensitivity of dark matter haloes to their accretion histories}

\author[M. P. Rey and A. Pontzen]
{Martin P. Rey$^{1}$\thanks{Contact e-mail: \href{martin.rey.16@ucl.ac.uk}{martin.rey.16@ucl.ac.uk}},
Andrew Pontzen$^{1}$, Am\'elie Saintonge$^{1}$
\\
$^{1}$Department of Physics and Astronomy, University College London, London WC1E 6BT, UK}

\date{23/10/2018}

\pubyear{2018}

\begin{document}
\label{firstpage}
\pagerange{\pageref{firstpage}--\pageref{lastpage}}
\maketitle

\begin{abstract}
  We apply our recently proposed ``quadratic genetic modification''
  approach to generating and testing the effects of alternative mass
  accretion histories for a single $\Lambda$CDM halo.  The goal of the
  technique is to construct different formation histories, varying the
  overall contribution of mergers to the fixed final mass.  This
  enables targeted studies of galaxy and dark matter halo formation's
  sensitivity to the smoothness of mass accretion.  Here, we focus on
  two dark matter haloes, each with four different mass accretion
  histories.  We find that the concentration of both haloes
  systematically decreases as their merger history becomes smoother.
  This causal trend tracks the known
  correlation between formation time and concentration parameters in
  the overall halo population. At fixed formation time, we further establish that halo
  concentrations are sensitive to the order in which mergers happen. This
  ability to study an individual halo's response to variations in its
  history is highly complementary to traditional methods based on
  emergent correlations from an extended halo population.
\end{abstract}




\section{Introduction} \label{section:intro}

In a $\Lambda$CDM universe, galaxies form and evolve embedded inside dark matter haloes.
The mass of the halo is believed to be
the primary driver of most galaxy properties; for example, empirical models of galaxy formation have often relied on
a parametrized mapping between dark matter halo mass and galaxy stellar mass (see
\citealt{Wechsler2018} for a review). Dark matter halo mass is also the main parameter of
halo clustering models used to recover cosmological
information from galaxy surveys (\citealt{Cooray2002}).

However, dark matter haloes
grow over time through hierarchical merging: smaller building-blocks merge together
to assemble larger haloes. The same mass at a given time
could have been assembled in many different ways:
through accretion of numerous small bodies or through a smaller
number of more significant events. This diversity of possible mass accretion histories
(hereafter MAHs) at fixed halo mass is thought to
generate scatter on both the galaxy-halo relationship (\citealt{Moster2013, RodriguezPuebla2016})
and in halo clustering bias (\citealt{Gao2005, Wechsler2006, Wetzel2007}).
The scatter can be further characterised by investigating the role of a halo's ``secondary'' properties
(e.g. density concentration, spin, or age). The evolution of these secondary
properties is shaped by
the response of dark matter haloes to external factors,
such as mergers and large-scale environment.

For a given halo, mergers and large-scale environment are seeded stochastically
from inflationary perturbations. This poses a challenge in studying any physical processes related
to secondary properties; the most common solution to date has been to
simulate large numbers of haloes to sample possible MAHs and cosmological environment at a given
mass scale (\citealt{Bullock2001b, Wechsler2002, Maccio2007, Ludlow2013, Klypin2016},
though see also \citealt{Zhao2003}).
Emergent correlations in the halo population have been characterised with such methods, e.g.
the relationship between halo concentration and mass (\citealt{Ludlow2014, Diemer2015, Klypin2016}).
But statistical sampling intrinsically makes it hard to construct
causal models because, by definition, every degree of freedom changes from one halo to the next.
For example, halo concentrations have been found to be correlated with
halo formation time (\citealt{Wechsler2002}),
halo environment (\citealt{AvilaReese2005, Maccio2007, Maulbetsch2007, Lee2017})
or halo spin (\citealt{Maccio2007}) but the interpretation of these correlations is
still under debate. These uncertainties propagate into empirical and semi-analytical models of
galaxy formation that rely on a physical account of the link between dark matter halo properties
and galactic properties.

Recently, genetic modification (hereafter GM, \citealt{Roth2016}) was introduced
as a method to study the response of a halo or galaxy to a controlled change in its merger
history. GMs create different versions of the
same halo, each with carefully specified modifications, while maintaining the same cosmological
large-scale structure.
One can then compare a range of scenarios for the formation of a particular halo or galaxy
in a simulation, keeping all degrees of freedom fixed except those specifically targeted.
The first application to galaxy formation was made by \citet{Pontzen2017}
who studied the response of a galaxy's star formation history to increased or decreased merger activity.

In practice, modifications are made in the initial conditions, and the simulation is then performed again.
Control over MAHs is achieved by modifying
the height and broadness of density peaks in the linear universe, motivated by
analytical structure formation theories (\citealt{Press1974, BBKS1986, Bond1991}).
This was achieved in \citet{Pontzen2017} by manually
tracking each merging substructure to the initial conditions and
modifying each region to obtain the required merger history. Tackling
multiple mergers with such a method risks spiralling complexity.
To simplify this procedure, \citet{Rey2018} presented
an extension to the existing framework targeting the local variance of the density field.
Variance encodes the height of multiple peaks and troughs compared to a mean value,
so should allow us to obtain direct control over the importance of multiple mergers.

The aim of this work is two-fold: to demonstrate that variance modifications from \citet{Rey2018}
provide the expected control over the overall smoothness of merger histories, and to show how GMs
can develop a causal account of the role of merger histories in shaping halo secondary properties.
We will show that a study of a small number of GM objects
complements existing large population studies. The paper is organized as follows.
In Section~\ref{section:setup}, we outline the procedure to generate
modified initial conditions and evolve them to $z=0$. In Section~\ref{section:families},
we construct specific families of modified haloes and
demonstrates that variance modifications directly control MAHs
at fixed mass in the way anticipated by \citet{Rey2018}.
In Section~\ref{section:halo}, we show that the continuous range of GM scenarios
causally links the details of MAHs to population-level variations in halo secondary properties.
We summarize our results
and conclude in Section~\ref{section:concl}.

\section{Numerical setup} \label{section:setup}

\begin{figure*}
  \centering
    \includegraphics[width=\textwidth]{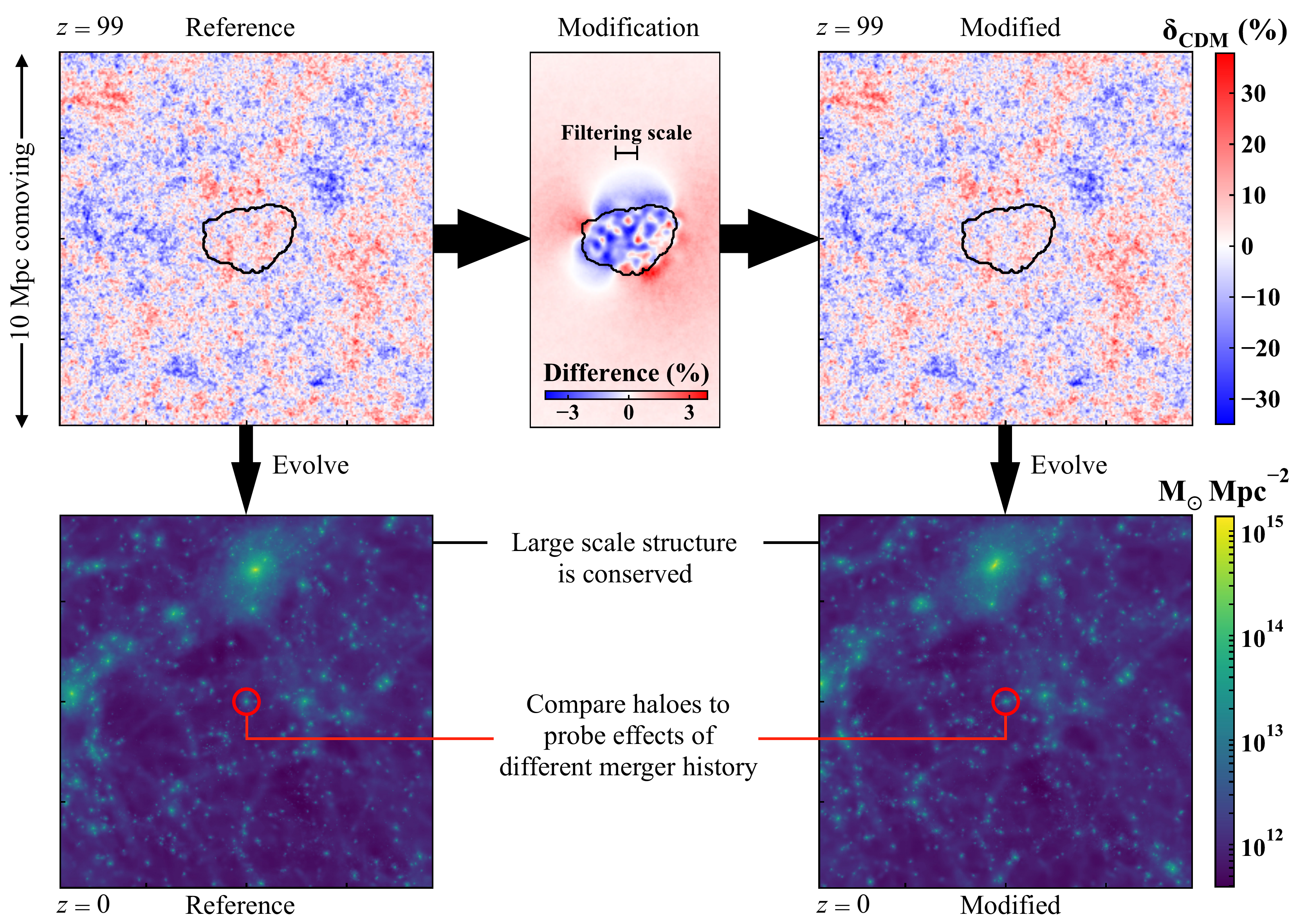}

    \caption{Workflow for generating a genetically modified halo from a reference halo.
    The top left panel shows a slice through the density field at $z=99$, i.e. the initial conditions
    for the reference halo. We evolve the density field to $z=0$ (bottom left)
    to obtain the properties of this halo, highlighted
    by a red circle of 3 virial radii. We then construct a
    genetically modified initial condition (top middle and right panels)
    with variance modifications (see Section~\ref{section:setup:ics}).
    We modify the smoothness of merger histories
    by redistributing the density structure inside the targeted region
    (black contour enclosing all particles tracked back from the reference halo at $z=0$).
    The large-scale structure (bottom right) matches that of the reference run (bottom left) but
    the merger histories of the two haloes are different
    (e.g. Fig.~\ref{fig:merger_reduction_839}) as well as their structural properties (e.g. Fig.~\ref{fig:nfw}).
    }
    \label{fig:setup}
\end{figure*}

We start by reviewing the necessary ingredients to generate a family of genetically
modified haloes. The theoretical underpinnings of GMs are described
in \citet{Roth2016} and \cite{Rey2018}; we focus here on the practical aspects of the procedure.
We describe how we define modifications in Section~\ref{section:setup:ics} and
in Section~\ref{section:setup:ramses} how we evolve initial conditions to $z=0$ in
a cosmological context.

\subsection{Initial conditions} \label{section:setup:ics}

In this section, we describe the general method of selecting haloes for re-simulation with the zoom
technique \citep{Katz1993} and generating a family of modified initial conditions using the quadratic
algorithm of \citet{Rey2018}.

To set up the zoom regions, we start from a simulation with uniform resolution
and select a halo, which will become the reference halo of the modified family. (The specific two haloes chosen for this work are described in Section \ref{section:families:modif}.)
We track back the centre of this halo to the initial condition and, for this work, open a spherical region
of radius comoving $3 \, \hMpc$ in which the mass resolution is refined. The
large extent of the zoom region is a simple way to ensure that the halo of interest is not
contaminated by heavy, i.e. low-resolution, particles in our present context where computing time is not a limiting factor. We verified that all haloes are
contaminated to less than $0.1$ per cent at $z=0$, including after modifications which can lead to different particles falling into a halo.
The initial conditions are defined from a flat $\Lambda \mathrm{CDM}$ cosmology (\citealt{Planck2015}),
with $h=0.6727$, $\Omega_{m} = 0.3139$, $\Omega_{b} = 0.04916$, $\Omega_{\Lambda} = 0.686095$,
$\sigma_{8} = 0.8440$, $n_{s}=0.9645$ and evolved using the Zel'dovich approximation
\citep{Zeldovich1970} to $z=99$.
All simulations have a box size of $50 \, \hMpc \approx 74 \, \Mpc$
and distances are stated in comoving units unless otherwise stated.

From these zoom initial conditions, we now generate modified
initial conditions. Our goal is to modify the merger history of a given
halo while keeping its final mass constant. Modifying the mean density in the Lagrangian region
of a halo has a direct impact on the final halo mass (see \citealt{Roth2016}), but we expect
variance modifications predominantly to redistribute mass inside the region,
hence acting on the merger history (see \citealt{Rey2018}).

Several ingredients are required to define these modifications:
\begin{enumerate}
  \item\label{item:window} the spatial windowing. We target the Lagrangian region of the unmodified halo
  by tracing back to $z=99$ all particles found in the halo at $z=0$. {\red{We define
  the boundary of a halo as the spherical radius, $\rvir$, at which the mass density is equal to
  $200$ times the critical density of the Universe.}}

  \item\label{item:filt} a mass scale for merging substructures. Variance modifications
  are designed to act on the peaks and troughs of the density field \textit{at a given scale}.
  To motivate the choice of the spatial filtering scale, we use the approximation that in the initial
  conditions, mass $M$ is linked to a scale $R$ via
  $M \approx 4 \pi \, \bar{\rho} \, R^{3}/3$, where $\bar{\rho}$ is the average
  density of the Universe. {\red{In practice, we choose the spatial scale corresponding to the mass at infall
  of the targeted merging substructures.}}
  Note that it is not necessary
  to identify specific substructures before generating modifications, unlike in the linear GM case.

  \item\label{item:linear} a control for the final halo mass. We adjust the mean {\red{overdensity}}
  inside the targeted region to fix the halo mass of the modified halo at $z=0$.

\end{enumerate}

Choosing the values of these generic parameters depends both on the halo in hand
and the requested modifications to its MAH. We will describe in Section~\ref{section:families:modif}
the details of the halo families used in this work.

\subsection{N-body evolution} \label{section:setup:ramses}

Once the reference and modified initial conditions have been generated, we need
to evolve them to $z=0$. All simulations presented in this work are dark matter only;
we use \textsc{ramses} \citep{Teyssier2002} which follows
N-body evolution with a particle-mesh method and cloud-in-cell interpolation.
The mesh on which the forces are calculated is adaptively refined over the course
of the simulation. We allow mesh refinement when 8 dark matter particles
are inside the same cell, up to a maximum spatial resolution of $1.1 \, \kpc$.
The mass of dark matter particles is $\xMsol{1.5}{7}$ and $\xMsol{1.2}{8}$
inside and outside the zoom regions, respectively.
We save 50 snapshots, equally spaced in scale factor between $z=99$ and $z=0$.

We identify dark matter haloes using the HOP halo finder as described in \citet{Eisentein1998},
and discard haloes with fewer than $100$ particles.
To calculate halo properties and merger trees, we make use of the \textsc{pynbody} and \textsc{tangos}
software packages (\citealt{Pontzen2013, Pontzen2018}). A key aspect of this work
is to compute the build-up of mass in a halo. \textsc{tangos} uses the unique ID
carried by each dark matter particle to match a halo with its successor in time, based
on the fraction of common particles between two structures. Repeating this procedure
for each snapshot constructs halo merger trees which are stored in a database.
In our \textsc{genetic} initial conditions generator and modifier, IDs are generated self-consistently between modified and unmodified simulations, allowing
us also to match haloes across different simulations.

We use the shrinking-sphere algorithm (\citealt{Power2003}) to determine halo centres.
At each timestep, we define the virial mass of a halo $\Mvir$ as the mass enclosed
within the spherical radius, $\rvir$ at the output redshift.

\section{Controlling the smoothness of merger histories} \label{section:families}

We now describe the specific modification setup used in this work.
Fig.~\ref{fig:setup} presents a graphical summary of the procedure described in
Section~\ref{section:setup}, with an example reference and variance-modified halo.
The top row focusses on the initial conditions of the two haloes,
while the bottom row shows the resulting evolution at $z=0$. We show slices of the overdensity
field for the reference (top left) and modified (top right) initial conditions, as well as a slice through
the difference between these fields (top middle). The two initial conditions are similar
except in the target region (black contour). The
filtering scale defines the scale at which density changes are targeted
inside the target region.

Leakage outside the targeted region can be observed in the difference
field. This is a feature of the GM algorithm: density perturbations
are correlated on all scales in a $\Lambda$CDM universe. Since the algorithm
is maintaining the correct power spectrum, it requires changes to have
some level of non-local impact. In this case, the leakage
is sufficiently minimal to visually recover
near-identical large-scale environments at $z=0$ (bottom panel),
and we therefore leave it unconstrained. We quantitatively discuss the impact of this choice
in Section~\ref{section:halo:env}.

\subsection{Creating modified haloes}\label{section:families:modif}

We now turn to the construction of two specific halo families that we will use in the remainder of this work.
Since the greatest number of applications will eventually be in hydrodynamical galaxy formation problems, we select two haloes
at the peak of star formation efficiency (\citealt{Behroozi2013}), i.e.
$\Mvir \approx 10^{12} \, \Msol$ at $z=0$. These two haloes, halo 740 and halo 839, are selected to have
different MAHs and environment, so that we can explore the effect of variance
modifications in different regimes. The modifications performed on these two haloes
are summarized in Table~\ref{table:runs}.

\begin{table}
  \centering
  \setlength\tabcolsep{4.0pt}
     \begin{tabular}{||c c c c||}
     \hline
     Simulation name & Mean changes (\%) & Variance changes (\%)\\
     \hline

     Volume-L50N512 & N/A & N/A\\

     Halo 740 family & $\{=, +10, =, =, =, -5\}$ &
     $\{=, +10, -5, -8, -10, -20\}$\\

     Halo 839 family & $\{=, =, =, =, =\}$
     & $\{=, -10, +10, +20, +30\}$\\

     \hline
     \end{tabular}

   \caption{Description of the modifications made in this work.
   From a uniform resolution volume, we select two target haloes (740 and 839)
   as reference for halo families. We create modified initial conditions from these reference
   haloes: changes in the mean density of the Lagrangian region control the halo mass at $z=0$,
   while changes in the
   variance control the smoothness of MAHs (see Section~\ref{section:families:modif}) around
   a targeted mass scale. Changes are quoted with respect to the value of the reference halo.
   }
   \label{table:runs}
\end{table}

Halo 839 has a quiet merger history with a series of small events
prior to $z\sim2$ and steady accretion of mass thereafter (see Fig.~\ref{fig:merger_reduction_839})
We generate a family of modified haloes targeting the merging structures around $z\sim2$,
enhancing their significance by following the procedure described in Section~\ref{section:setup:ics}.
The merging structures have a mass of~$\sim \xMsol{1.5}{10}$ {\red{at first infall}}, motivating a
filtering scale of $0.30 \, \hMpc$. To enhance merger significance, we increase
incrementally the variance by 10, 20 and 30 per cent relative to the value in the reference run.
We also decrease the variance by 10 per cent to explore whether the merger history can be made
even smoother. We emphasize that each member of the family sits in a large-scale
environment minimally modified from the reference halo
and {\red{that conserving the mean density in the Lagrangian region ensures that
halo masses match at $z = 0$ to within 5 per cent.}}

Halo 740 has a more complex merger history, dominated by two equal-mass
mergers in less than $1 \, \Gyr$ around $z\sim1$ (see Fig.~\ref{fig:merger_reduction_740}).
We wish to make the overall history smoother by decreasing the local variance.
The merging structures around $z\sim1$ have masses~$\sim \xMsol{2}{11}$, motivating a
filtering scale of $0.70 \, \hMpc$. We again adopt an incremental approach,
decreasing the variance by 5, 10 and 20 per cent relative to the value in the reference run.
We also increase the variance by 10 per cent to explore if the merger history can be made
even rougher.
{\red{The first initial runs were performed conserving the mean overdensity in the Lagrangian region.
While motivated by analytical structure formation, the mean overdensity is a property
of the linear density field, and hence not a perfect predictor of the strongly non-linear halo mass.
Following these initial modifications, we observed variations in the final halo masses
of halo 740's family members of up to 10 per cent. These larger changes compared to halo 839
are due to modifications targeting larger structures,
closer in scale to the overall final halo mass.}}
For complete clarity we modified the mean density as stated in Table~\ref{table:runs} and
re-ran the simulations, finding agreement improved to match within 5 per cent.

We emphasize that the strength of the GM method lies in its incremental approach.
By making continuous changes to the initial conditions and studying the consequent non-linear
response of the halo, we can pinpoint the tipping points of halo and galaxy formation, where
small changes in merger histories have large consequences on observed properties.
We study next the detailed impact of our modifications on the MAHs and
merger tree structure of both haloes (Section~\ref{section:gf}), as well as how these changes are reflected
in dark matter halo properties (Section~\ref{section:halo}).

\subsection{Results} \label{section:gf}

\begin{figure}
  \centering
    \includegraphics[width=\columnwidth]{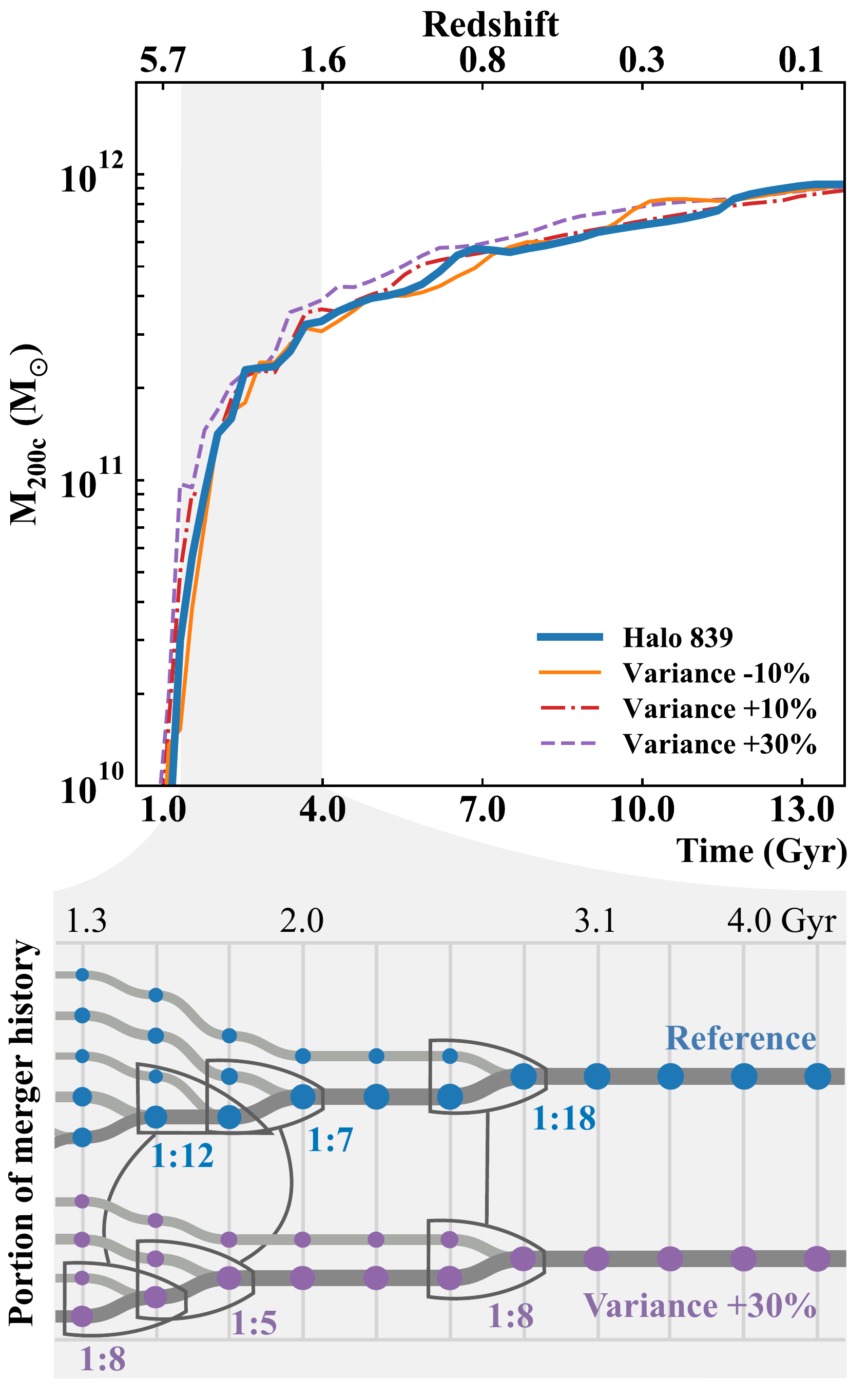}

    \caption{The growth of virial mass over cosmic time (top panel) for the reference (blue thick curve)
    halo 839 and its modified counterparts. Halo 839 has a quiet merger history with its
    most significant activity being three minor mergers around $z \sim 2$.
    The impact of variance modifications
    is hence most visible on the merger trees (bottom panel). We show
    the reference halo (blue) and the modified halo with increased variance by 30 per cent (purple),
    with highlighted mergers matched between the trees.
    Increasing the variance successfully increases the mass ratios of all three mergers.
    }
    \label{fig:merger_reduction_839}
\end{figure}

Fig.~\ref{fig:merger_reduction_839} presents the results of variance modifications
on halo 839. The top panel shows the MAHs over cosmic time of the reference halo
(thick blue) and three family members with variance
decreased by 10 per cent and increased by 10 and 30 per cent (orange solid, red dotted and purple dashed respectively).
Variance modifications targeted several
merger events around $z\sim2$, highlighted in time
by the grey band. We show in the bottom panel the merger trees
of the reference halo and the modified halo with increased variance by 30 per cent in this time window.
The size of branches scales logarithmically with their mass;
the darker bottom branch in each case is the major progenitor.
Merging events are matched between simulations
(see Section~\ref{section:setup:ramses}) and highlighted by the linked black boxes.
We quantify merger mass ratios by the ratio between the number of particles
inside haloes as found by the halo finder. We have verified that changing
to merger ratios as defined with $\Mvir$
or {\red{the mass within one scale radius}} \citep{Hopkins2010} does not impact our conclusions.

By inspecting the merger trees, we can confirm the ability of a single
variance modification to control multiple mergers.
The reference halo 839 has three minor events with merger ratios
1:12, 1:7 and 1:18. Increasing the variance by 30 per cent increases the merger ratios
of all three mergers, respectively to 1:8, 1:5 and 1:8. Thus, we have successfully
made the targeted mergers of halo 839 more significant, confirming
 the viability of the quadratic modification approach to
controlling MAHs.

\begin{figure}
  \centering
    \includegraphics[width=\columnwidth]{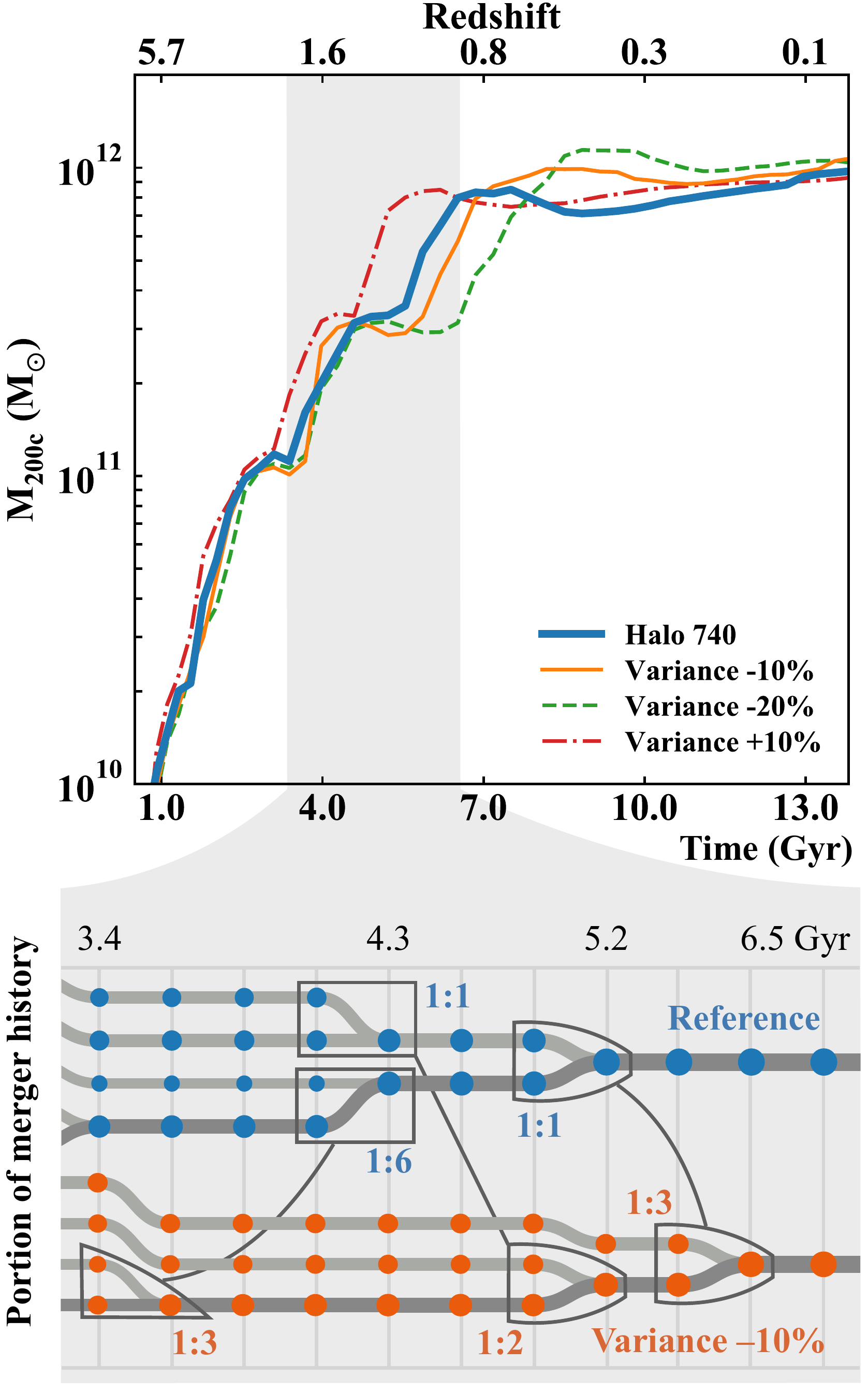}

    \caption{As Fig.~\ref{fig:merger_reduction_839} but for halo 740,
      which has a more complex merger history that is dominated by
      equal-mass mergers around $z=1$. Decreasing the variance by 10
      per cent successfully smooths the impact of these two mergers,
      as measured by their respective merger ratios (orange
      tree). Because we fix the halo mass at $z=0$, there must be a
      compensating increase in mass from other sources: in fact a 1:6
      merger in the reference run (blue tree) increases significance
      to 1:3. Thus, one minor and two major mergers in the reference
      simulation are mapped onto three medium mergers in the modified
      case.  This also generates a change in the topology of the
      trees: unlike in the reference, all three structures directly
      merge into the major progenitor in the modified case.  }
    \label{fig:merger_reduction_740}
\end{figure}

Fig.~\ref{fig:merger_reduction_740} presents the results of variance modifications
to halo 740. As before, we show the MAHs of the reference halo and three selected family members
(top panel),
as well as the merger trees for the reference halo and the modified halo with
variance decreased by 10 per cent (bottom panels).
The reference halo 740 has three major events in the window of interest:
two roughly equal-mass mergers (ratios 1:1) that we explicitly
targeted with our choice of filtering scale and a less significant 1:6 event.
Reducing the variance by 10 per cent reduces the merger ratios of the two targeted
mergers (now 1:2 and 1:3). However, the required compensation in mass accretion
results in increasing the merger ratio of the smaller
event to 1:3. Since we fix the halo mass at $z=0$, modifications
inevitably redistribute the mass between substructures
to obtain convergence of MAHs at late time.
The minimal nature of GM naturally ensures that MAHs converge before
the targeted $z\sim1$ mergers.

Variance modifications for halo 740 create a reconfiguration of the merger tree topology.
Comparing the merger trees for the reference halo and the $-10$ per cent modified halo
(bottom panel of Fig.~\ref{fig:merger_reduction_740}), we
find that the same substructures are incorporated into the main progenitor
(darker bottom branch of each tree) following two different patterns:
\begin{enumerate}
\renewcommand{\theenumi}{(\Alph{enumi})}
  \item\label{item:A} In the reference merger tree (blue panel) the top two branches
  merge together.
  The merger remnant from the top two branches is later incorporated into the main progenitor.
  Over the course of the time window, the main progenitor experiences \textit{two} mergers.

  \item\label{item:B} In the $-10$ per cent variance merger tree (orange panel), the same top
  two branches remain independent. They are incorporated turn by turn into
  the main progenitor, which now experiences \textit{three} mergers over the course of the time window.

\end{enumerate}

Our incremental approach, with continuous changes to the initial conditions,
allows us to explore the merger tree configuration as a function of local variance. For halo 740,
the reference halo, as well as $+10$ and $-5$ per cent modified haloes
follow scenario \ref{item:A},
while $-10$ and $-20$ per cent follow scenario \ref{item:B}.
Note that the reconfiguration tipping point is a property of this particular halo; our variance modifications on halo 839 never modify the topology of the merger tree, for example.
We will see in Section~\ref{section:halo}
that merger tree topology has consequences for the evolution of
dark matter halo properties that cannot be discerned from population studies alone.

In addition to merger ratios and tree structure,
variance modifications also impact the timing of mergers. A systematic outcome can be observed
for both families: reducing merger ratios pushes mergers to later times; conversely, increasing
ratios pushes mergers earlier. This effect was also observed in \citet{Pontzen2017}
using a different modification setup.
In fact, it is a generic outcome of any GM procedure, since
the linear density field is
given by the divergence of the velocity
field in $\Lambda$CDM initial conditions.
Reducing merger ratios is achieved by smoothing density gradients, in turn
reducing the relative initial velocities of the two substructures.
These two structures will then take longer to coalesce, effectively pushing the
merger to later time. If fixing the timing of a merger is paramount,
one can construct and impose a new velocity
modification to conserve the peculiar velocity structure inside the Lagrangian
region, at the cost of increasing leakage effects outside the region. We leave the study of
simulations with such additional modifications to future work.

In summary, variance modifications create different versions of the reference halo with:
\begin{enumerate}
  \item a minimally modified large-scale structure (see Fig.~\ref{fig:setup}), which
  we will quantify in Section~\ref{section:halo:env};
  \item a predictable but non-trivial effect on MAHs: increased variance increases the mass
  ratios of mergers and shifts them to slightly earlier times, and may reconfigure the tree
  topology when critical thresholds are exceeded
  (see Fig.~\ref{fig:merger_reduction_839} and~\ref{fig:merger_reduction_740});
  \item a minimal impact on the remaining MAH, which
  converge before and after the area targeted by variance modifications
  (e.g. Fig.~\ref{fig:merger_reduction_740}).
\end{enumerate}

The combination of these features shows that we have achieved the objective of simulating
essentially the same halo, up to the selected modifications. Variance modifications create a
``dial'' through which the MAH of a given halo can be seen as a tuneable parameter. This opens the door
for a wide variety of applications, both for galaxy formation studies
and for dark matter halo physics on which we will focus next.

\section{Response of haloes to changes in their merger history} \label{section:halo}

We demonstrated in Section~\ref{section:gf} our ability to generate different versions
of the same halo with systematically altered accretion history.
We now study the response of $z=0$ dark matter halo properties to these
varying MAHs. Since our families of modified haloes have been generated with fixed final halo mass,
we focus on variations in halo secondary properties such as
halo concentration, formation time or spin. The statistical correlations between
these properties have been extensively studied
(e.g. \citealt{Bullock2001b, Wechsler2002, Maccio2007, Ludlow2014, Klypin2016}).
However, the physical drivers of the emergent correlations and their scatter
remains uncertain due to the extremely large number of degrees of freedom in the initial conditions.
Genetic modifications alter a small subset of the degrees of freedom, allowing us to
gain new physical insight into secondary properties.

We compare our families of modified haloes to a large statistical sample
extracted from the Volume-L50N512 simulation, our
highest uniform resolution simulation (see Table~\ref{table:runs}) with~$\sim 38000$ haloes at $z=0$.
Comparing to the overall halo population will allow us to quantify the relative impact of modifications
with respect to population-level variations.

\subsection{Calculating halo secondary properties}
The properties that we measure are as follows.

\begin{enumerate}
  \item \textit{Concentration}: We calculate halo concentrations, $\cnfw$,
  using the velocity profile method (described in \citealt{Prada2012,Klypin2016}).
  This method is based on computing the ratio of peak circular velocity
  to circular velocity at $\rvir$ (see Section \ref{section:setup:ramses}) as a measure of halo concentration.
  This concentration is a profile-independent quantity so is not
  impacted by the goodness of fit to a specific analytical form
  (e.g. \citealt{NFW1997}, hereafter NFW, or \citealt{Einasto1965}). To obtain the velocity ratio,
  we compute the enclosed mass profile in bins evenly spaced in log radius between $0.7\,\kpc$
  and $\rvir$. From this, we calculate the circular
  velocity profile from which we can obtain the desired velocity ratio.
  We remap this concentration measure into the well-known NFW concentration
  using equation 20 of \citet{Klypin2016} to facilitate comparison with past
  studies. We have checked that our results are unchanged if
  using the velocity ratio as a direct measure of concentration rather than
  remapping to the NFW definition.

  \item\textit{Spin}: We calculate halo spin, $\spin$,
  following the definition of \citet{Bullock2001a}, equation 5, inside a sphere of radius $\rvir$.

  \item \textit{Formation redshift}: {\red{We define the formation redshift, $\zform$,
  as the redshift at which the main progenitor has accreted the mass enclosed within
  the NFW scale radius at $z=0$ (\citealt{Ludlow2014, Correa2015}).}}

  \item \textit{Environment}: We define halo environment, $\env$,
  by calculating the density inside a sphere of radius $10 \, \Mpc$
  centred on the halo. We divide by the mean matter
  density of the Universe to obtain a dimensionless quantity following \citet{Maccio2007}.
\end{enumerate}

Our goal is to study how halo secondary properties vary following variance modifications.
There are, however, potential sources of purely numerical variations,
which we investigate briefly before turning to the main results.
The first is the impact of numerical resolution
(see e.g. \citealt{Power2003}
for halo density profiles). Using the same volume simulation
but with degraded mass resolution of $\xMsol{9.6}{8}$,
we compare the median and 68 per cent confidence intervals of all secondary properties in the mass
range $\xMsol{5}{11} < \Mvir < \xMsol{5}{12}$ at these two resolutions.
We find they are different by less than 4 per cent for halo concentrations and less
than 1 per cent for other secondary properties and therefore conclude that our results
will not be sensitive to resolution. Second, large variations in halo secondary
properties can also be caused by propagation of numerical noise in an intrinsically chaotic system,
i.e. the ``butterfly effect'' (\citealt{Keller2018, Genel2018}).
To exclude this possibility, we ran all simulations in the family of halo 740 three
times. {\red{Our simulation code, \textsc{ramses}, does not conserve the order of arithmetic operations,
hence ensuring that roundoff errors are seeded differently between each re-run.}}
We find that rerunning yields variations
in halo secondary properties to up to 2 per cent.  We will see next that both variations
remain small compared to the overall effect of genetic modifications
and we hence conclude that our results
are robust to numerical issues.

\subsection{The evolution of concentration and formation time} \label{section:halo:cM}

\begin{figure*}
  \centering
    \includegraphics[width=\textwidth]{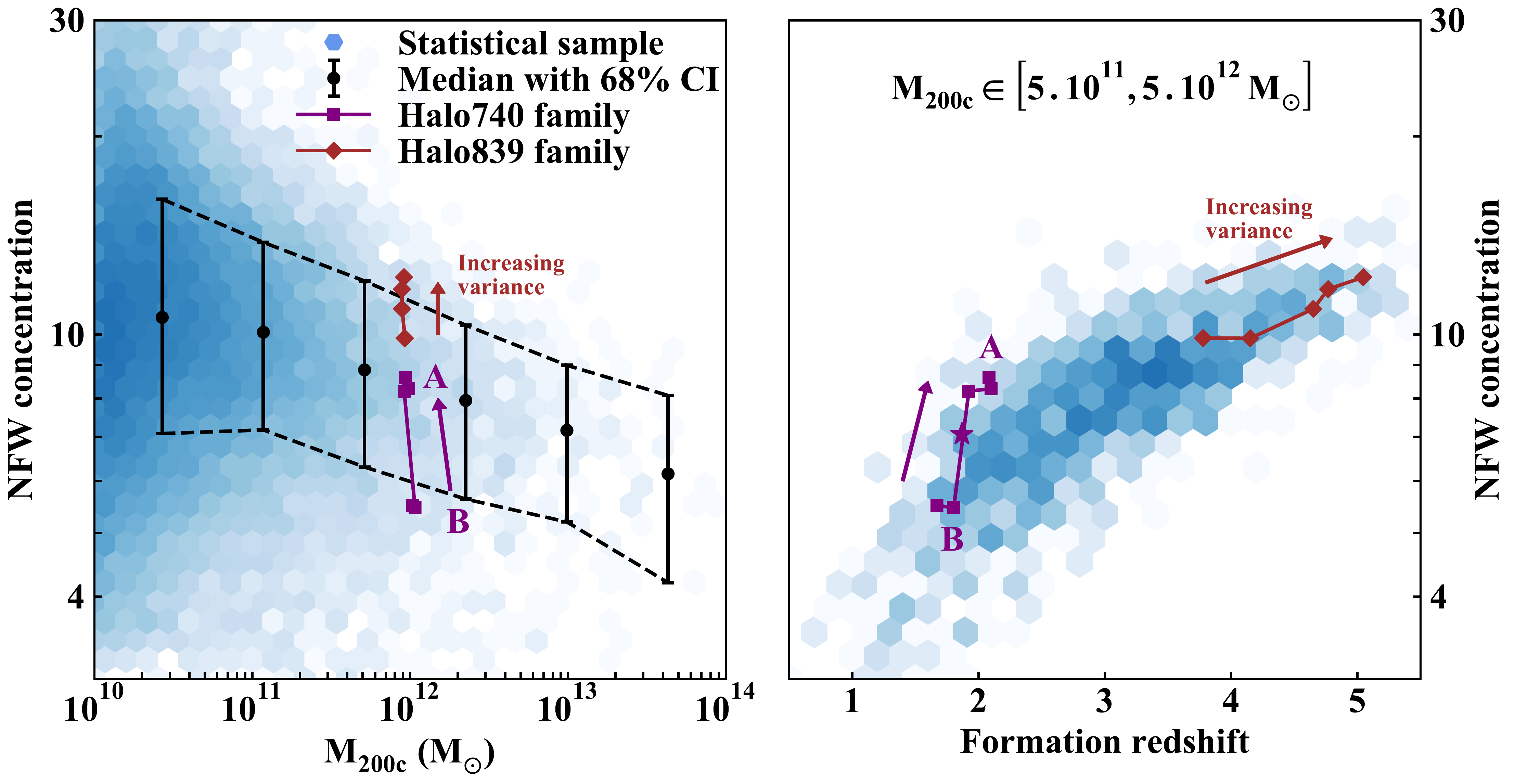}
    \caption{
    The concentration-mass relation at $z=0$ (left) for the two modified families (halo 740
    and halo 839 in purple squares and red diamonds respectively).
    Increasing the variance (arrows) in the initial conditions systematically yields more concentrated
    haloes. We compare these variations with correlations in the overall halo population (blue hexagon bins
    with median and 68 per cent confidence intervals in six bins of halo mass).
    Variations in the mass accretion history of two haloes generate changes in concentration
    that are comparable to the entire population-level scatter.
    These variations are consistent with the correlation between halo
    concentration and formation redshift (right), as both families move along
    the direction of the correlation. The jump in halo concentration
    between \ref{item:A} and~\ref{item:B} is tied to the
    reconfiguration of the merger tree shown in Fig.~\ref{fig:merger_reduction_740}.
    {\red{To verify this, we constructed an additional simulation (purple star), intermediate between the ends of the jump.
    We found that this version undergoes a three-way merger (i.e. it sits precisely
    on the transition between the two merger topologies) and as a result its
    concentration is also intermediate.}}
    }
    \label{fig:nfw}
\end{figure*}

Fig.~\ref{fig:nfw} presents halo properties for the two families
of modified haloes (purple squares and red diamonds for halo 740 and halo 839
respectively) compared to the halo statistical sample at $z=0$ (blue hexagon bins).
Each modified halo has been engineered to have systematically varied
MAH at fixed halo mass.
Arrows show the direction of increasing variance for each family.

The left-hand panel of Fig.~\ref{fig:nfw} shows the concentration-mass relation at $z=0$. Our halo population
recovers the well-known trend that the median NFW concentration (black points) decreases with
increased halo mass, although with considerable scatter (\citealt{Bullock2001b, Ludlow2014, Klypin2016}).
Physically, this scatter is thought to be generated in part by the variety of
MAHs at given halo mass (\citealt{Wechsler2002, Ludlow2013}).
We cleanly demonstrate this causal link here:
systematically varying MAH yields a systematic change in concentration at fixed halo mass.

{\red{
The link between MAHs and concentration has been extensively studied through
empirical correlations (\citealt{Wechsler2002, Zhao2003, Gao2005, Gao2008}) and analytical models
(\citealt{Ludlow2014, Correa2015, Ludlow2016}). The strongest
predictions are obtained by relating the early, fast mass accretion phase (summarised by $\zform$) to the
final halo concentration.}}
At fixed mass, a halo assembling its central mass earlier
(when the Universe was denser) will be more
centrally concentrated. The population in our volume reproduces this trend
(Fig.~\ref{fig:nfw}, right-hand panel) in the mass interval $\xMsol{5}{11} < \Mvir < \xMsol{5}{12}$.
We see from both families that increasing variance also incrementally increases
the formation redshift of modified haloes. Thus the population-level correlation is
reproduced as a causal connection in the family-level studies.

{\red{The evolution of both families is gradual along the direction of the correlation: increasing
the variance of a given halo makes it form earlier, in turn making its concentration higher.
This is especially visible in the case of halo 839 in which our modifications targeted early
mergers, hence smoothly modifying its formation redshift.
By way of contrast, the halo 740 family shows less variation in formation redshift $\zform$
because modifications targeted late-time mergers, leaving the early history unchanged.
Nonetheless, we observe a significant variation in halo concentration, highlighted by a}} ``jump'' of $\Delta \cnfw = 2.6$
in halo concentration from $\cnfw = 5.5$ to $8.1$, untied to any significant variation
in formation redshift. This discontinuity is significant at the population level, in the
sense that $\Delta \cnfw =  2.2 \, \sigma$, where $\sigma$ is defined as half the $68$ per cent confidence interval of the population.
It occurs at the point that the merger tree is reconfigured, as
described in Section~\ref{section:gf}. All members of the family
with a merger tree structure following~\ref{item:A} (i.e. with one combined merger into the major progenitor) are found on the high end of the
discontinuity in concentration, while
all members following~\ref{item:B} (with two mergers into the major progenitor) are on the low end.
No such reconfigurations
are ever generated in halo 839's family and similarly, we do not observe a discontinuous response
in halo concentration.

Our incremental approach with variance modifications
therefore allows us to tie the origin of this concentration jump
to the merger tree reconfiguration. We can explain \textit{a posteriori} why the
two different scenarios generate vastly different concentrations.
Scenario~\ref{item:A} is dominated
by an equal-mass merger on a mostly radial orbit. The mass being brought
by this penetrates deeper in the potential well of the main progenitor,
leading to a higher concentration. By contrast, the same mass in scenario~\ref{item:B}
is incorporated smoothly through two merger events with smaller
mass ratios. The in-falling mass is more evenly distributed through the final halo,
leading to an overall lower concentration.

{\red{One of the key feature of the GM approach is its ability to refine around tipping points
by generating additional intermediate scenarios.
To further test our explanation that the merger tree reconfiguration is the source of the concentration jump, we generate
a new modification from halo 740 with variance decreased by 8 per cent (in between
the two ends of the jump which have variance decreased by respectively 5 per cent and 10 per cent). In this intermediate case,
the previously described mergers combine into a three-way event,
i.e. the transition scenario in merger tree topology between~\ref{item:A} and~\ref{item:B}.
The purple star in Fig.~\ref{fig:nfw} shows the resulting
formation time and concentration of this halo which, as expected, bisects the two ends of the jump.
This confirms the causal connection between the merger scenario and the concentration.}}

We conclude that the memory of the different merger configurations between scenario~\ref{item:A} and~\ref{item:B}
is retained to $z=0$ and generates a discontinuity in the resulting halo concentrations.
We emphasize that this discontinuity is not stochastic, but rather the result of mapping continuous,
incremental changes in the initial conditions to a discrete merger tree topology (i.e. mergers happen in this order or do not).
{\red{The discrete transition is due to major mergers happening later than the halo's formation time
and hence cannot be captured by models mapping the early mass assembly onto halo concentrations.
New summary statistics beyond formation time}} would be needed to encapsulate knowledge about merger tree topology,
for example by counting the number of mergers weighted by their mass ratios. We leave
an exploration of such new summary measures as future work and now turn our
attention to other secondary halo properties and their relationships with halo concentration.

\begin{figure*}
  \centering
  \includegraphics[width=\textwidth]{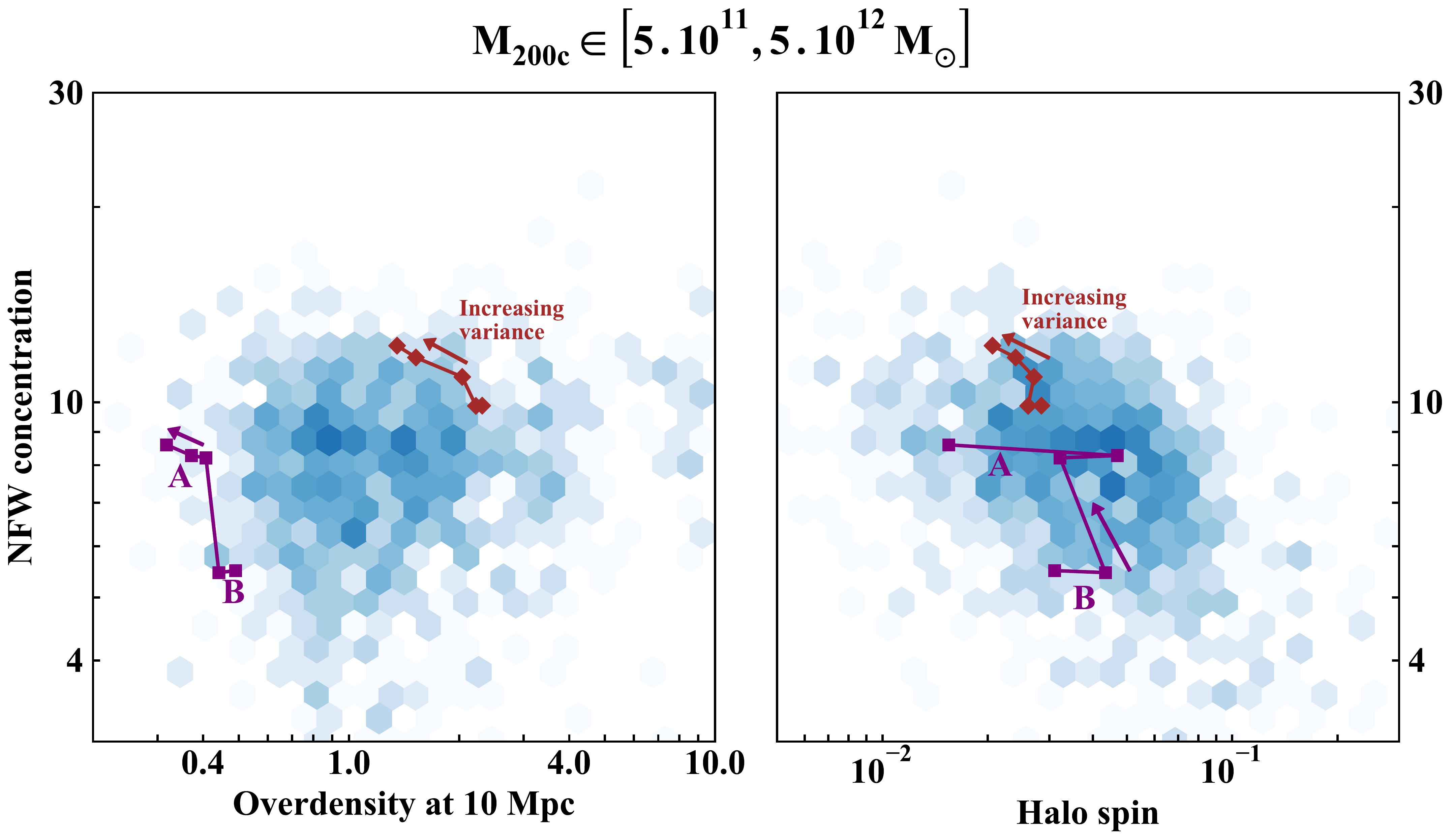}
  \caption{The impact of variance modifications on the local
    environment of haloes (left) and halo spin (right). Increasing the
    variance generates modified haloes in slightly underdense
    environments compared to their previous version, because internal
    variance is correlated with a halo's external density in a
    $\Lambda$CDM universe.  We recover a weak correlation between halo
    spin and halo concentration. Halo 740 shows a jump in spin
    between its two highest variance scenarios, which can be explained
    by a late-time minor merger which has been pushed to $z<0$
    (i.e. into the future) in the highest variance, lowest-spin case. }
  \label{fig:c_env}
\end{figure*}

\subsection{Halo environment} \label{section:halo:env}

Another potential source of scatter in the concentration-mass relationship is the diversity
of halo local environments at fixed mass
(\citealt{AvilaReese2005, Maccio2007, Maulbetsch2007, Lee2017}), especially at low halo mass.
We now investigate the response of halo environments in families of modified haloes.
Visually the large scale structure is unaltered (Fig.~\ref{fig:setup}), but we should consider a more quantitative measure.

The left panel of Fig.~\ref{fig:c_env} shows such a measure of environment, the local overdensity at
10 $\Mpc$ against halo concentration. The mass bin from which the halo population
is extracted and the colour coding are the same as Fig.~\ref{fig:nfw}. We recover no significant
correlation between this measure of environment and halo concentration at this mass
in the halo population, consistent with past studies (e.g. \citealt{Maccio2007, Lee2017}).

We observe a significant and systematic variation with increasing variance in both families,
compared to the population scatter, with increasing variance
systematically pushes haloes to underdense environment. More specifically,
halo 740 and halo 839's entire families go from $\env = 0.42$ to $0.31$ and $\env = 2.31$ to $1.35$
respectively, with increasing variance. This evolution can be compared
to the population, ranging $0.66 \, \sigma$ and $0.81 \, \sigma$ respectively where we have defined $\sigma$
on the log of the population since cosmological densities are log-normally distributed (e.g. \citealt{Coles1991}).

Given the visual similarity of the large scale structures in a family of simulations (Fig.~\ref{fig:setup}),
these differences should be explained.
The GM algorithm by construction minimises the differences
between reference and modified regions, but also seeks to maintain the correlation function
of the field.
In the case of variance modifications, we create
changes on small scales (i.e. internal to the halo) which then leak to larger scales
due to these intrinsic correlations
(see the top centre panel of Fig.~\ref{fig:setup} for a visual example of leakage).
Decreasing the variance on small scales means decreasing the density contrasts
in this region, hence creating an \textit{overdensity} compared to the reference halo
as a compensation effect. This directly translates into a
higher environment density as observed in Fig.~\ref{fig:c_env}.

In other words, the drift in environment is a result of the algorithm seeking to maintain
maximally likely surroundings for the particular field realisation inside the target halo region.
{\red{Given that environments correlate very weakly with halo concentrations and spin at this mass scale,
we do not try to correct for this drift. Nonetheless,}}
if a conserved environmental density
is paramount, this can be accomplished with further stipulations on
the surrounding field. {\red{For example, one may add an additional modification explicitly maintaining constant
the mean overdensity of the environment on a chosen scale.}}
In the limit of a perfectly conserved environment, each pixel outside the
Lagrangian region could be fixed to its reference value. Though technically feasible,
such a drastic approach might result in an unlikely draw from the $\Lambda$CDM power spectrum
as all degrees of freedom in the field would be constrained. We will investigate
the strengths and drawbacks of such an approach in future work.

\subsection{Halo spin} \label{section:halo:spin}

Finally, we explore the effect of variance modifications
on angular momentum.  Halo spin and concentration have a weak but significant
correlation (\citealt{Maccio2007}) and spin is often used as proxy for galactic
properties, especially for disc formation (see \citealt{Somerville2008, Benson2012},
though see also \citealt{Jiang2018}).

The right-hand panel of Fig.~\ref{fig:c_env} shows the evolution
of halo spin at $z=0$ and concentration for both halo families. Tracks are compared
to a halo population extracted from the same mass bin as in the left-hand panel.
A weak correlation is observed in the halo population, similar to \citet{Maccio2007}
but remains tentative due the low number of haloes in our simulation volume.

Halo 839's gradual evolution seems to follow the direction of the correlation, but halo 740
presents a more chaotic behaviour. The previously reported discontinuity in halo concentration
between scenarios \ref{item:A} and \ref{item:B} is only linked to a change
from $\spin = 0.032$ to $0.043$, i.e. $\Delta \spin = 0.48 \,\sigma$ where $\sigma$
is defined on the log of the population since halo spin are log-normally distributed (\citealt{Bullock2001b}).
However, another discontinuity is visible in halo spin, previously invisible in halo concentration,
between the two highest variance points of halo 740 (the reference halo and the $+10$ per cent halo). This discontinuity
ranges from $\spin = 0.047$ to $0.015$ with increasing variance,
corresponding to a significance of $\Delta \spin = 1.8 \,\sigma$ when compared
to the overall spin population.

Unlike for halo concentrations, this jump in halo spin is not tied to a reconfiguration
of the merger tree topology. Halo spin has been observed to peak
around merger times (\citealt{Vitvitska2002}), as merging bodies bring in fresh
angular momentum during in-fall. Halo spin then decays as smooth halo evolution proceeds and
the in-falling body is destroyed.
In the reference run, a small merger (mass ratio 1:12) happens at $z=0.2$, making
halo spin peak but without enough time to decay before the end of the simulation.
In the +10 per cent halo, the same merger has not yet been incorporated
in the main progenitor by the end
of the simulation at $z=0$, hence not generating the same increase in halo spin
and yielding an overall much lower $z=0$ spin value.

The fact that a minor merger can generate a variation in spin comparable to the
breadth of the entire population
highlights the sensitivity of the halo spin to the detailed merger history.
The amplitude and direction of spins, the mass ratio and
the orbital in-fall of the two merging bodies all likely play a role; we leave
a more detailed analysis of their interplay as future work.

\section{Conclusion}\label{section:concl}

We have demonstrated our ability to construct alternative versions of a
cosmological dark matter halo
with fixed final halo mass and varying smoothness for the mass accretion history.
This turns merger histories into a tuneable parameter, opening a new route for galaxy
formation and dark matter halo numerical studies.

Control over the smoothness of the mass accretion history is achieved by modifying
the local variance on a given scale inside the Lagrangian region of a halo,
while maintaining the Gaussianity and the ensemble power spectrum. We implemented the
algorithm presented in \citet{Rey2018} for simultaneous linear and quadratic
modifications, and applied it successfully in the context of cosmological
zoom simulations (see Fig.~\ref{fig:setup}).

We demonstrated our ability to successfully reduce or increase the mass ratios
of multiple mergers (see Fig.~\ref{fig:merger_reduction_839})
by respectively reducing or enhancing the variance, achieving direct control
of the overall smoothness of MAHs at fixed final halo mass. The implementation
allows us to target mergers of a given mass scale, making minimal impact on the
remaining MAH and large-scale structure around the halo.

The strength of this framework lies in its incremental approach. By gradually
varying the local variance, we can causally explore the internal response of
dark matter haloes. Specifically, we targeted ``tipping points'' for which
a small change in MAH has large consequences on observed properties. In this way,
we established that a reconfiguration in the merger tree topology was responsible for generating
a large jump in halo concentrations (see Fig.~\ref{fig:merger_reduction_740} and~\ref{fig:nfw}),
comparable to the overall population-level scatter.
This kind of dramatic sensitivity to initial conditions may be responsible for much
of the scatter in halo and galaxy properties. The sensitivity is not equivalent to
stochasticity arising due to chaotic amplification of numerical inaccuracies \citep{Keller2018, Genel2018},
which we explicitly ruled out as playing any part in our results. Rather, it
arises naturally from the mapping of continuous initial conditions
to a discrete set of mergers. GMs are therefore
highly complementary to population-level studies for understanding how
correlations and their scatter emerge through the complexity of halo formation.

In addition to dark matter halo physics, the method will prove invaluable to
investigate the role of merger histories in shaping properties of the galaxy population.
Mergers can, for example, directly impact star-formation rates by fuelling starbursts
(\citealt{Springel2005}) or alter galaxy morphology (\citealt{diMatteo05,Naab2009,Johansson09,Pontzen2017}).
Additionally, memory
of past mergers is likely retained in the stellar kinematics in the outskirts of galaxies (i.e. stellar haloes),
due to the collisionless nature and long orbital time-scales of such stars.
The newly discovered diversity (\citealt{Merritt2016, Monachesi2016}) in stellar haloes
seems to be generated by merger histories with varying smoothness
(\citealt{Elias2018, Monachesi2018}),
as well as through the response of the central galaxy to these mergers
(\citealt{Zolotov2009, Cooper2015}).
A handful of variance modified galaxies could help disentangling how merger histories
build a stellar halo, as well as clarifying the observational imprints of these mergers.

\section*{Acknowledgements}
We thank Andrey Kravtsov, Justin Read and Michael Tremmel for helpful discussions on the
interpretation of the results and comments on an earlier version of the manuscript.
We also thank the anonymous referee for a detailed report that helped improve our paper.
MR acknowledges support
from the Perren Fund and the IMPACT fund. AP is supported by the
Royal Society. AS acknowledges support from the Royal Society
through the award of a University Research Fellowship.



\bibliographystyle{mnras}
\bibliography{Biblio}

\begin{thebibliography}{}
\makeatletter
\relax
\def\mn@urlcharsother{\let\do\@makeother \do\$\do\&\do\#\do\^\do\_\do\%\do\~}
\def\mn@doi{\begingroup\mn@urlcharsother \@ifnextchar [ {\mn@doi@}
  {\mn@doi@[]}}
\def\mn@doi@[#1]#2{\def\@tempa{#1}\ifx\@tempa\@empty \href
  {http://dx.doi.org/#2} {doi:#2}\else \href {http://dx.doi.org/#2} {#1}\fi
  \endgroup}
\def\mn@eprint#1#2{\mn@eprint@#1:#2::\@nil}
\def\mn@eprint@arXiv#1{\href {http://arxiv.org/abs/#1} {{\tt arXiv:#1}}}
\def\mn@eprint@dblp#1{\href {http://dblp.uni-trier.de/rec/bibtex/#1.xml}
  {dblp:#1}}
\def\mn@eprint@#1:#2:#3:#4\@nil{\def\@tempa {#1}\def\@tempb {#2}\def\@tempc
  {#3}\ifx \@tempc \@empty \let \@tempc \@tempb \let \@tempb \@tempa \fi \ifx
  \@tempb \@empty \def\@tempb {arXiv}\fi \@ifundefined
  {mn@eprint@\@tempb}{\@tempb:\@tempc}{\expandafter \expandafter \csname
  mn@eprint@\@tempb\endcsname \expandafter{\@tempc}}}

\bibitem[\protect\citeauthoryear{{Avila-Reese}, {Col{\'{\i}}n},
  {Gottl{\"o}ber}, {Firmani}  \& {Maulbetsch}}{{Avila-Reese}
  et~al.}{2005}]{AvilaReese2005}
{Avila-Reese} V.,  {Col{\'{\i}}n} P.,  {Gottl{\"o}ber} S.,  {Firmani} C.,
  {Maulbetsch} C.,  2005, \mn@doi [\apj] {10.1086/491726}, 634, 51

\bibitem[\protect\citeauthoryear{{Bardeen}, {Bond}, {Kaiser}  \&
  {Szalay}}{{Bardeen} et~al.}{1986}]{BBKS1986}
{Bardeen} J.~M.,  {Bond} J.~R.,  {Kaiser} N.,   {Szalay} A.~S.,  1986, \mn@doi
  [\apj] {10.1086/164143}, 304, 15

\bibitem[\protect\citeauthoryear{{Behroozi}, {Wechsler}  \&
  {Conroy}}{{Behroozi} et~al.}{2013}]{Behroozi2013}
{Behroozi} P.~S.,  {Wechsler} R.~H.,   {Conroy} C.,  2013, \mn@doi [\apj]
  {10.1088/0004-637X/770/1/57}, 770, 57

\bibitem[\protect\citeauthoryear{{Benson}}{{Benson}}{2012}]{Benson2012}
{Benson} A.~J.,  2012, \mn@doi [\na] {10.1016/j.newast.2011.07.004}, 17, 175

\bibitem[\protect\citeauthoryear{{Bond}, {Cole}, {Efstathiou}  \&
  {Kaiser}}{{Bond} et~al.}{1991}]{Bond1991}
{Bond} J.~R.,  {Cole} S.,  {Efstathiou} G.,   {Kaiser} N.,  1991, \mn@doi
  [\apj] {10.1086/170520}, 379, 440

\bibitem[\protect\citeauthoryear{{Bullock}, {Kolatt}, {Sigad}, {Somerville},
  {Kravtsov}, {Klypin}, {Primack}  \& {Dekel}}{{Bullock}
  et~al.}{2001a}]{Bullock2001b}
{Bullock} J.~S.,  {Kolatt} T.~S.,  {Sigad} Y.,  {Somerville} R.~S.,  {Kravtsov}
  A.~V.,  {Klypin} A.~A.,  {Primack} J.~R.,   {Dekel} A.,  2001a, \mn@doi
  [\mnras] {10.1046/j.1365-8711.2001.04068.x}, 321, 559

\bibitem[\protect\citeauthoryear{{Bullock}, {Dekel}, {Kolatt}, {Kravtsov},
  {Klypin}, {Porciani}  \& {Primack}}{{Bullock} et~al.}{2001b}]{Bullock2001a}
{Bullock} J.~S.,  {Dekel} A.,  {Kolatt} T.~S.,  {Kravtsov} A.~V.,  {Klypin}
  A.~A.,  {Porciani} C.,   {Primack} J.~R.,  2001b, \mn@doi [\apj]
  {10.1086/321477}, 555, 240

\bibitem[\protect\citeauthoryear{{Coles} \& {Jones}}{{Coles} \&
  {Jones}}{1991}]{Coles1991}
{Coles} P.,  {Jones} B.,  1991, \mn@doi [\mnras] {10.1093/mnras/248.1.1}, 248,
  1

\bibitem[\protect\citeauthoryear{{Cooper}, {Parry}, {Lowing}, {Cole}  \&
  {Frenk}}{{Cooper} et~al.}{2015}]{Cooper2015}
{Cooper} A.~P.,  {Parry} O.~H.,  {Lowing} B.,  {Cole} S.,   {Frenk} C.,  2015,
  \mn@doi [\mnras] {10.1093/mnras/stv2057}, 454, 3185

\bibitem[\protect\citeauthoryear{{Cooray} \& {Sheth}}{{Cooray} \&
  {Sheth}}{2002}]{Cooray2002}
{Cooray} A.,  {Sheth} R.,  2002, \mn@doi [\physrep]
  {10.1016/S0370-1573(02)00276-4}, 372, 1

\bibitem[\protect\citeauthoryear{{Correa}, {Wyithe}, {Schaye}  \&
  {Duffy}}{{Correa} et~al.}{2015}]{Correa2015}
{Correa} C.~A.,  {Wyithe} J. S.~B.,  {Schaye} J.,   {Duffy} A.~R.,  2015,
  \mn@doi [\mnras] {10.1093/mnras/stv1363}, 452, 1217

\bibitem[\protect\citeauthoryear{{Di Matteo}, {Springel}  \& {Hernquist}}{{Di
  Matteo} et~al.}{2005}]{diMatteo05}
{Di Matteo} T.,  {Springel} V.,   {Hernquist} L.,  2005, \mn@doi [\nat]
  {10.1038/nature03335}, 433, 604

\bibitem[\protect\citeauthoryear{{Diemer} \& {Kravtsov}}{{Diemer} \&
  {Kravtsov}}{2015}]{Diemer2015}
{Diemer} B.,  {Kravtsov} A.~V.,  2015, \mn@doi [\apj]
  {10.1088/0004-637X/799/1/108}, 799, 108

\bibitem[\protect\citeauthoryear{{Einasto}}{{Einasto}}{1965}]{Einasto1965}
{Einasto} J.,  1965, Trudy Astrofizicheskogo Instituta Alma-Ata, 5, 87

\bibitem[\protect\citeauthoryear{{Eisenstein} \& {Hut}}{{Eisenstein} \&
  {Hut}}{1998}]{Eisentein1998}
{Eisenstein} D.~J.,  {Hut} P.,  1998, \mn@doi [\apj] {10.1086/305535}, 498, 137

\bibitem[\protect\citeauthoryear{{Elias}, {Sales}, {Creasey}, {Cooper},
  {Bullock}, {Rich}  \& {Hernquist}}{{Elias} et~al.}{2018}]{Elias2018}
{Elias} L.~M.,  {Sales} L.~V.,  {Creasey} P.,  {Cooper} M.~C.,  {Bullock}
  J.~S.,  {Rich} R.~M.,   {Hernquist} L.,  2018, \mn@doi [\mnras]
  {10.1093/mnras/sty1718}, 479, 4004

\bibitem[\protect\citeauthoryear{{Gao}, {Springel}  \& {White}}{{Gao}
  et~al.}{2005}]{Gao2005}
{Gao} L.,  {Springel} V.,   {White} S.~D.~M.,  2005, \mn@doi [\mnras]
  {10.1111/j.1745-3933.2005.00084.x}, 363, L66

\bibitem[\protect\citeauthoryear{{Gao}, {Navarro}, {Cole}, {Frenk}, {White},
  {Springel}, {Jenkins}  \& {Neto}}{{Gao} et~al.}{2008}]{Gao2008}
{Gao} L.,  {Navarro} J.~F.,  {Cole} S.,  {Frenk} C.~S.,  {White} S. D.~M.,
  {Springel} V.,  {Jenkins} A.,   {Neto} A.~F.,  2008, \mn@doi [\mnras]
  {10.1111/j.1365-2966.2008.13277.x}, 387, 536

\bibitem[\protect\citeauthoryear{{Genel} et~al.,}{{Genel}
  et~al.}{2018}]{Genel2018}
{Genel} S.,  et~al., 2018, preprint (\mn@eprint {arXiv} {1807.07084})

\bibitem[\protect\citeauthoryear{{Hopkins} et~al.,}{{Hopkins}
  et~al.}{2010}]{Hopkins2010}
{Hopkins} P.~F.,  et~al., 2010, \mn@doi [\apj] {10.1088/0004-637X/724/2/915},
  724, 915

\bibitem[\protect\citeauthoryear{{Jiang} et~al.,}{{Jiang}
  et~al.}{2018}]{Jiang2018}
{Jiang} F.,  et~al., 2018, preprint (\mn@eprint {arXiv} {1804.07306})

\bibitem[\protect\citeauthoryear{{Johansson}, {Naab}  \& {Burkert}}{{Johansson}
  et~al.}{2009}]{Johansson09}
{Johansson} P.~H.,  {Naab} T.,   {Burkert} A.,  2009, \mn@doi [\apj]
  {10.1088/0004-637X/690/1/802}, 690, 802

\bibitem[\protect\citeauthoryear{{Katz} \& {White}}{{Katz} \&
  {White}}{1993}]{Katz1993}
{Katz} N.,  {White} S.~D.~M.,  1993, \mn@doi [\apj] {10.1086/172935}, 412, 455

\bibitem[\protect\citeauthoryear{{Keller}, {Wadsley}, {Wang}  \&
  {Kruijssen}}{{Keller} et~al.}{2018}]{Keller2018}
{Keller} B.~W.,  {Wadsley} J.~W.,  {Wang} L.,   {Kruijssen} J.~M.~D.,  2018,
  preprint (\mn@eprint {arXiv} {1803.05445})

\bibitem[\protect\citeauthoryear{{Klypin}, {Yepes}, {Gottl{\"o}ber}, {Prada}
  \& {He{\ss}}}{{Klypin} et~al.}{2016}]{Klypin2016}
{Klypin} A.,  {Yepes} G.,  {Gottl{\"o}ber} S.,  {Prada} F.,   {He{\ss}} S.,
  2016, \mn@doi [\mnras] {10.1093/mnras/stw248}, 457, 4340

\bibitem[\protect\citeauthoryear{{Lee}, {Primack}, {Behroozi},
  {Rodr{\'{\i}}guez-Puebla}, {Hellinger}  \& {Dekel}}{{Lee}
  et~al.}{2017}]{Lee2017}
{Lee} C.~T.,  {Primack} J.~R.,  {Behroozi} P.,  {Rodr{\'{\i}}guez-Puebla} A.,
  {Hellinger} D.,   {Dekel} A.,  2017, \mn@doi [\mnras]
  {10.1093/mnras/stw3348}, 466, 3834

\bibitem[\protect\citeauthoryear{{Ludlow} et~al.,}{{Ludlow}
  et~al.}{2013}]{Ludlow2013}
{Ludlow} A.~D.,  et~al., 2013, \mn@doi [\mnras] {10.1093/mnras/stt526}, 432,
  1103

\bibitem[\protect\citeauthoryear{{Ludlow}, {Navarro}, {Angulo},
  {Boylan-Kolchin}, {Springel}, {Frenk}  \& {White}}{{Ludlow}
  et~al.}{2014}]{Ludlow2014}
{Ludlow} A.~D.,  {Navarro} J.~F.,  {Angulo} R.~E.,  {Boylan-Kolchin} M.,
  {Springel} V.,  {Frenk} C.,   {White} S.~D.~M.,  2014, \mn@doi [\mnras]
  {10.1093/mnras/stu483}, 441, 378

\bibitem[\protect\citeauthoryear{{Ludlow}, {Bose}, {Angulo}, {Wang},
  {Hellwing}, {Navarro}, {Cole}  \& {Frenk}}{{Ludlow}
  et~al.}{2016}]{Ludlow2016}
{Ludlow} A.~D.,  {Bose} S.,  {Angulo} R.~E.,  {Wang} L.,  {Hellwing} W.~A.,
  {Navarro} J.~F.,  {Cole} S.,   {Frenk} C.~S.,  2016, \mn@doi [\mnras]
  {10.1093/mnras/stw1046}, 460, 1214

\bibitem[\protect\citeauthoryear{{Macci{\`o}}, {Dutton}, {van den Bosch},
  {Moore}, {Potter}  \& {Stadel}}{{Macci{\`o}} et~al.}{2007}]{Maccio2007}
{Macci{\`o}} A.~V.,  {Dutton} A.~A.,  {van den Bosch} F.~C.,  {Moore} B.,
  {Potter} D.,   {Stadel} J.,  2007, \mn@doi [\mnras]
  {10.1111/j.1365-2966.2007.11720.x}, 378, 55

\bibitem[\protect\citeauthoryear{{Maulbetsch}, {Avila-Reese}, {Col{\'{\i}}n},
  {Gottl{\"o}ber}, {Khalatyan}  \& {Steinmetz}}{{Maulbetsch}
  et~al.}{2007}]{Maulbetsch2007}
{Maulbetsch} C.,  {Avila-Reese} V.,  {Col{\'{\i}}n} P.,  {Gottl{\"o}ber} S.,
  {Khalatyan} A.,   {Steinmetz} M.,  2007, \mn@doi [\apj] {10.1086/509706},
  654, 53

\bibitem[\protect\citeauthoryear{{Merritt}, {van Dokkum}, {Abraham}  \&
  {Zhang}}{{Merritt} et~al.}{2016}]{Merritt2016}
{Merritt} A.,  {van Dokkum} P.,  {Abraham} R.,   {Zhang} J.,  2016, \mn@doi
  [\apj] {10.3847/0004-637X/830/2/62}, 830, 62

\bibitem[\protect\citeauthoryear{{Monachesi}, {Bell}, {Radburn-Smith},
  {Bailin}, {de Jong}, {Holwerda}, {Streich}  \& {Silverstein}}{{Monachesi}
  et~al.}{2016}]{Monachesi2016}
{Monachesi} A.,  {Bell} E.~F.,  {Radburn-Smith} D.~J.,  {Bailin} J.,  {de Jong}
  R.~S.,  {Holwerda} B.,  {Streich} D.,   {Silverstein} G.,  2016, \mn@doi
  [\mnras] {10.1093/mnras/stv2987}, 457, 1419

\bibitem[\protect\citeauthoryear{{Monachesi} et~al.,}{{Monachesi}
  et~al.}{2018}]{Monachesi2018}
{Monachesi} A.,  et~al., 2018, preprint (\mn@eprint {arXiv} {1804.07798})

\bibitem[\protect\citeauthoryear{{Moster}, {Naab}  \& {White}}{{Moster}
  et~al.}{2013}]{Moster2013}
{Moster} B.~P.,  {Naab} T.,   {White} S.~D.~M.,  2013, \mn@doi [\mnras]
  {10.1093/mnras/sts261}, 428, 3121

\bibitem[\protect\citeauthoryear{{Naab}, {Johansson}  \& {Ostriker}}{{Naab}
  et~al.}{2009}]{Naab2009}
{Naab} T.,  {Johansson} P.~H.,   {Ostriker} J.~P.,  2009, \mn@doi [\apjl]
  {10.1088/0004-637X/699/2/L178}, 699, L178

\bibitem[\protect\citeauthoryear{{Navarro}, {Frenk}  \& {White}}{{Navarro}
  et~al.}{1997}]{NFW1997}
{Navarro} J.~F.,  {Frenk} C.~S.,   {White} S.~D.~M.,  1997, \mn@doi [\apj]
  {10.1086/304888}, 490, 493

\bibitem[\protect\citeauthoryear{{Planck Collaboration} et~al.,}{{Planck
  Collaboration} et~al.}{2016}]{Planck2015}
{Planck Collaboration} et~al., 2016, \mn@doi [\aap]
  {10.1051/0004-6361/201525830}, 594, A13

\bibitem[\protect\citeauthoryear{{Pontzen} \& {Tremmel}}{{Pontzen} \&
  {Tremmel}}{2018}]{Pontzen2018}
{Pontzen} A.,  {Tremmel} M.,  2018, \mn@doi [The Astrophysical Journal
  Supplement Series] {10.3847/1538-4365/aac832}, 237, 23

\bibitem[\protect\citeauthoryear{{Pontzen}, {Ro{\v s}kar}, {Stinson}  \&
  {Woods}}{{Pontzen} et~al.}{2013}]{Pontzen2013}
{Pontzen} A.,  {Ro{\v s}kar} R.,  {Stinson} G.,   {Woods} R.,  2013, {pynbody:
  N-Body/SPH analysis for python}, Astrophysics Source Code Library (\mn@eprint
  {ascl} {1305.002})

\bibitem[\protect\citeauthoryear{{Pontzen}, {Tremmel}, {Roth}, {Peiris},
  {Saintonge}, {Volonteri}, {Quinn}  \& {Governato}}{{Pontzen}
  et~al.}{2017}]{Pontzen2017}
{Pontzen} A.,  {Tremmel} M.,  {Roth} N.,  {Peiris} H.~V.,  {Saintonge} A.,
  {Volonteri} M.,  {Quinn} T.,   {Governato} F.,  2017, \mn@doi [\mnras]
  {10.1093/mnras/stw2627}, 465, 547

\bibitem[\protect\citeauthoryear{{Power}, {Navarro}, {Jenkins}, {Frenk},
  {White}, {Springel}, {Stadel}  \& {Quinn}}{{Power} et~al.}{2003}]{Power2003}
{Power} C.,  {Navarro} J.~F.,  {Jenkins} A.,  {Frenk} C.~S.,  {White} S.~D.~M.,
   {Springel} V.,  {Stadel} J.,   {Quinn} T.,  2003, \mn@doi [\mnras]
  {10.1046/j.1365-8711.2003.05925.x}, 338, 14

\bibitem[\protect\citeauthoryear{{Prada}, {Klypin}, {Cuesta}, {Betancort-Rijo}
  \& {Primack}}{{Prada} et~al.}{2012}]{Prada2012}
{Prada} F.,  {Klypin} A.~A.,  {Cuesta} A.~J.,  {Betancort-Rijo} J.~E.,
  {Primack} J.,  2012, \mn@doi [\mnras] {10.1111/j.1365-2966.2012.21007.x},
  423, 3018

\bibitem[\protect\citeauthoryear{{Press} \& {Schechter}}{{Press} \&
  {Schechter}}{1974}]{Press1974}
{Press} W.~H.,  {Schechter} P.,  1974, \mn@doi [\apj] {10.1086/152650}, 187,
  425

\bibitem[\protect\citeauthoryear{{Rey} \& {Pontzen}}{{Rey} \&
  {Pontzen}}{2018}]{Rey2018}
{Rey} M.~P.,  {Pontzen} A.,  2018, \mn@doi [\mnras] {10.1093/mnras/stx2744},
  474, 45

\bibitem[\protect\citeauthoryear{{Rodr{\'{\i}}guez-Puebla}, {Primack},
  {Behroozi}  \& {Faber}}{{Rodr{\'{\i}}guez-Puebla}
  et~al.}{2016}]{RodriguezPuebla2016}
{Rodr{\'{\i}}guez-Puebla} A.,  {Primack} J.~R.,  {Behroozi} P.,   {Faber}
  S.~M.,  2016, \mn@doi [\mnras] {10.1093/mnras/stv2513}, 455, 2592

\bibitem[\protect\citeauthoryear{{Roth}, {Pontzen}  \& {Peiris}}{{Roth}
  et~al.}{2016}]{Roth2016}
{Roth} N.,  {Pontzen} A.,   {Peiris} H.~V.,  2016, \mn@doi [\mnras]
  {10.1093/mnras/stv2375}, 455, 974

\bibitem[\protect\citeauthoryear{{Somerville}, {Hopkins}, {Cox}, {Robertson}
  \& {Hernquist}}{{Somerville} et~al.}{2008}]{Somerville2008}
{Somerville} R.~S.,  {Hopkins} P.~F.,  {Cox} T.~J.,  {Robertson} B.~E.,
  {Hernquist} L.,  2008, \mn@doi [\mnras] {10.1111/j.1365-2966.2008.13805.x},
  391, 481

\bibitem[\protect\citeauthoryear{{Springel}, {Di Matteo}  \&
  {Hernquist}}{{Springel} et~al.}{2005}]{Springel2005}
{Springel} V.,  {Di Matteo} T.,   {Hernquist} L.,  2005, \mn@doi [\mnras]
  {10.1111/j.1365-2966.2005.09238.x}, 361, 776

\bibitem[\protect\citeauthoryear{{Teyssier}}{{Teyssier}}{2002}]{Teyssier2002}
{Teyssier} R.,  2002, \mn@doi [\aap] {10.1051/0004-6361:20011817}, 385, 337

\bibitem[\protect\citeauthoryear{{Vitvitska}, {Klypin}, {Kravtsov}, {Wechsler},
  {Primack}  \& {Bullock}}{{Vitvitska} et~al.}{2002}]{Vitvitska2002}
{Vitvitska} M.,  {Klypin} A.~A.,  {Kravtsov} A.~V.,  {Wechsler} R.~H.,
  {Primack} J.~R.,   {Bullock} J.~S.,  2002, \mn@doi [\apj] {10.1086/344361},
  581, 799

\bibitem[\protect\citeauthoryear{{Wechsler} \& {Tinker}}{{Wechsler} \&
  {Tinker}}{2018}]{Wechsler2018}
{Wechsler} R.~H.,  {Tinker} J.~L.,  2018, preprint (\mn@eprint {arXiv}
  {1804.03097})

\bibitem[\protect\citeauthoryear{{Wechsler}, {Bullock}, {Primack}, {Kravtsov}
  \& {Dekel}}{{Wechsler} et~al.}{2002}]{Wechsler2002}
{Wechsler} R.~H.,  {Bullock} J.~S.,  {Primack} J.~R.,  {Kravtsov} A.~V.,
  {Dekel} A.,  2002, \mn@doi [\apj] {10.1086/338765}, 568, 52

\bibitem[\protect\citeauthoryear{{Wechsler}, {Zentner}, {Bullock}, {Kravtsov}
  \& {Allgood}}{{Wechsler} et~al.}{2006}]{Wechsler2006}
{Wechsler} R.~H.,  {Zentner} A.~R.,  {Bullock} J.~S.,  {Kravtsov} A.~V.,
  {Allgood} B.,  2006, \mn@doi [\apj] {10.1086/507120}, 652, 71

\bibitem[\protect\citeauthoryear{{Wetzel}, {Cohn}, {White}, {Holz}  \&
  {Warren}}{{Wetzel} et~al.}{2007}]{Wetzel2007}
{Wetzel} A.~R.,  {Cohn} J.~D.,  {White} M.,  {Holz} D.~E.,   {Warren} M.~S.,
  2007, \mn@doi [\apj] {10.1086/510444}, 656, 139

\bibitem[\protect\citeauthoryear{{Zel'dovich}}{{Zel'dovich}}{1970}]{Zeldovich1970}
{Zel'dovich} Y.~B.,  1970, \aap, 5, 84

\bibitem[\protect\citeauthoryear{{Zhao}, {Mo}, {Jing}  \& {B{\"o}rner}}{{Zhao}
  et~al.}{2003}]{Zhao2003}
{Zhao} D.~H.,  {Mo} H.~J.,  {Jing} Y.~P.,   {B{\"o}rner} G.,  2003, \mn@doi
  [\mnras] {10.1046/j.1365-8711.2003.06135.x}, 339, 12

\bibitem[\protect\citeauthoryear{{Zolotov}, {Willman}, {Brooks}, {Governato},
  {Brook}, {Hogg}, {Quinn}  \& {Stinson}}{{Zolotov} et~al.}{2009}]{Zolotov2009}
{Zolotov} A.,  {Willman} B.,  {Brooks} A.~M.,  {Governato} F.,  {Brook} C.~B.,
  {Hogg} D.~W.,  {Quinn} T.,   {Stinson} G.,  2009, \mn@doi [\apj]
  {10.1088/0004-637X/702/2/1058}, 702, 1058

\makeatother
\end{thebibliography}


\bsp	
\label{lastpage}
\end{document}